# 基于改进图注意力网络的电力系统关键环节辨识

王长刚[1,2]，王先伟[2]，曹宇[1,2]，李扬[1,2]，吕琪[2]，张耀心[3]

(1. 现代电力系统仿真控制与绿色电能新技术教育部重点实验室(东北电力大学)，吉林 吉林 132012；
2. 东北电力大学，吉林 吉林 132012；3. 国网天津市电力公司，天津 300010)

**摘要**：随着电网的扩大与新能源比例的增加，使得电网的不确定性和随机性因素增加危及系统运行安全，寻找出电网中的关键环节来保障电网运行时的可靠性就显得尤为重要。针对当前传统电网关键环节辨别方法识别速度慢、难以满足电网实际运行要求的问题，本文提出了基于改进图注意力网络算法(Modified Graph Attention Network, MGAT)的电网关键环节辨识方法。首先，结合复杂网络理论和电网实际运行数据建立评价指标集。之后，利用 MGAT 挖掘出电网运行时的各项指标与关键环节之间的映射关系，建立关键环节辨识模型，考虑到训练准确性和效率等需求，对原始数据进行预处理以优化模型。其次，通过仿真得到原始数据集，对辨识模型进行训练、验证和测试。最后，利用所述模型应用于改进的 IEEE 30 节点系统和实际电网中，结果表明本文方法具有可行性，且准确性和速度优于传统方法，有一定的工程利用价值。
**关键词**：关键环节辨识；复杂网络理论；图注意力神经网络；运行可靠性

**Critical link identification of power system vulnerability based on modified graph attention network**

WANG Changgang[1,2], WANG Xianwei[2], CAO Yu[1,2], LI Yang[1,2], LÜ Qi[2], ZHANG Yaoxin[3]

(1. Key Laboratory of Modern Power System Simulation and Control and Renewable Energy Technology, Ministry of Education, Northeast Electric Power University, Jilin 132012, China; 2. Northeast Electric Power University, Jilin 132012, China; 3. State Grid Tianjin Electric Power Company, Tianjin 300010, China)

**Abstract:** As the power grid expands and the proportion of renewable energy increases, the uncertainty and randomness of the power grid have increased, posing a threat to the security of system operation. It becomes crucial to identify the crucial links in the power grid to ensure its reliability during operation. Addressing the issues of slow identification speed and difficulty in meeting the actual operational requirements of traditional methods for identifying crucial links in the power grid, this paper proposes a method based on the Modified Graph Attention Network (MGAT) algorithm for identifying crucial link in the power grid. Firstly, combining complex network theory with actual power grid operational data, an evaluation index set is established. Subsequently, the MGAT is utilized to explore the mapping relationships between various indicators during power grid operation and the crucial links, forming a model for crucial link identification. Considering the requirements of training accuracy and efficiency, the original data is preprocessed to optimize the model. Next, through simulation, the original dataset is obtained for training, validation, and testing of the identification model. Finally, the proposed model is applied to an improved IEEE 30-node system and a real power grid. The results indicate the feasibility of the proposed method, with accuracy and speed surpassing traditional methods, demonstrating its practical engineering value.

## 0 引言

随着电网的逐步建设、各地区互联互通和新能源发电的兴起，电网的复杂度和不确定性增加，确定电网各个环节安全运行是保障电网可靠运行的前置条件[1-4]。在实际运行中，某些线路或节点在运行时起到牵一发而动全身的作用，如果这些环节发生故障，对电网安全运行会产生巨大影响，如何快速而准确的找到这些关键环节保证电力系统的运行可靠性是亟待解决的问题[5-6]。

目前，国内外对于电网关键环节的研究多包涵在电力系统脆弱性的研究之下，主要分为基于拓扑关系和运行状态两大类[8-12]。前者利用复杂网络理论选取特定的特征量（例如度数、介数），考虑实际

电网的拓扑结构，融入系统物理特性，构造脆弱性指标，进而评估出关键环节。后者是通过电网的实际运行状态，如潮流分布，来实现关键环节的辨识。文献[13]利用随机矩阵理论及熵理论对电网运行状态进行分析，提出了具有普适性的电网薄弱节点判断指标；文献[14]将复杂网络理论和运行状态有机结合，利用组合权重法对不同节点指标进行权重分配，建立电网薄弱点组合指标，并用实例证明有效性。文献[15]提出了基于结构保持模型的暂态能量函数法结合有向电气介数的支路暂态脆弱性综合评估方法，该方法可以快速判断容易失稳的脆弱支路以及临界机群，保证了电网安全稳定运行。文献[16]提出了考虑不同类型节点与相邻节点间传输转移的信息值及节点间非等概率传输特性的改进PageRank算法进行关键节点的辨识。这些方法都因计算繁琐耗时，不能很好地适应如今变换快速的电力系统，也难以实现在线辨识的要求。

随着近几年人工智能技术的兴起和发展，其中包括的深度学习算法在电力系统的应用也越广泛[17]。"离线训练，在线辨识"是深度学习众多强大的功能之一，能够快速准确的完成分类，预测等任务[18-21]。利用深度学习挖掘特征与标签之间的映射关系，建立分类模型是深度学习最为基本的任务之一。对于电力系统来说，它是一个由众多线路和节点所组成的庞大系统，拓扑结构是其相当重要的属性之一，传统深度学习在电力系统的应用只注意到挖掘数据之间的关系，没有注意到电力系统各部分之间的拓扑关系，进而导致模型的泛化能力不高。图深度神经网络[22]（Graph Neural Network，GNN）相较于其他的深度学习，不仅仅只关注自身的属性，还会通过拓扑关系聚合相邻节点的特征信息，使得特征提取过程更加立体，涵盖更多的信息。其在电力系统优化调度[23-25]方面、故障诊断[26-27]方面得到良好的应用。Veličković 等人[28]在此基础上提出了图注意力网络（Graph Attention Network，GAT），GAT 的节点在聚合邻居表示时采用了注意力机制，根据节点与邻居的相关性来为邻居分配注意力系数，最后通过节点邻居特征及节点自身的特征加权平均值来更新节点特征。Brody 等人[29]发现原始GAT 的注意力系数的排序系数在图中的所有节点之间共享，并且不受查询节点的限制。这一事实影响了 GAT 的表现力。其通过修改注意力系数公式，提高 GAT 的表达能力。即 MGAT(Modified Graph Attention Network)。

本文在现有的基础上，构建了基于改进图注意力网络的电力系统关键环节的辨识器。首先，通过改进复杂网络理论模型寻找出电力系统中的关键环节并建立样本集，其次，通过构建的样本集，利用MGAT 挖掘出电力系统运行时的各项数据和关键环节之间的映射关系，并构建辨识器。最后将本文方法应用于改进的 IEEE 30 节点系统和实际电网中，结果表明本文方法可行，且具有比传统计算方法更加快速、准确性的特点。

## 1 关键环节的定义

### 1.1 复杂网络理论模型

复杂网络理论是一种通过对网络拓扑结构特征量统计规律的研究来揭示各种复杂网络中共性的网络随机演化性质、结构稳定性及受到攻击时的动力学传播特性的理论方法[30]。其多适用于具有拓扑结构属性的对象，如关系网，分子结构等。复杂网络理论的特定参数指标如下：

（1）度数。指节点所连边的个数。很明显，度数越高，这个节点在整个图中就越重要。

（2）最短路径。网络之中两节点之间的距离可以用连接这两个节点的最短路径来表示，即在网络中需经过最少的边数。

（3）聚类系数。节点的聚类系数可以反应一个节点与其相邻节点的连接情况，与邻居节点之间实际相连的边数，比上可能存在最多相连边数的比值，便是聚类系数。

（4）介数。介数分为节点介数和边介数，定义节点或边的介数为在网络中所有的最短路径中，最短路径经过的某节点或边的路径数与总数的比值。

本文选取文献[31]的改进复杂网络理论模型作为电网关键环节评估准则。其不单仅仅只看节点或线路在整个拓扑上的重要性，而且还会考虑节点或线路在整个电网运行时各项数值。

### 1.2 关键环节的定义

电网中的关键环节分为关键节点和关键支路，电网中的某个节点的关键程度由优化节点度数簇和优化节点介数簇加权求和得到，具体公式如下[32]：

$$d_{Li} = \sum_{j}\left|\alpha_{ij}\right| \tag{1}$$

$$d_{Ci} = \max(\sum_{j,a_{ij}=1} c_{ij}\alpha_{ij}, \sum_{j,a_{ij}=-1} c_{ij}\left|\alpha_{ij}\right|) \tag{2}$$

$$b_{Hi} = \frac{\sum_{j\in G, k\in L; j\neq k}\sum_{m=1}^{M(jk)} n^{i}(m)}{\sum_{j\in G, k\in L; j\neq k} M(jk)} \tag{3}$$

$$b_{Li} = \frac{\sum_{j\in G,k\in L;j\neq k}\sum_{m=1}^{M(jk)} w_{jk}(m)d_{jk}(m)n^i(m)}{\sum_{j\in G,k\in L;j\neq k}\sum_{m=1}^{M(jk)} w_{jk}(m)d_{jk}(m)} \quad (4)$$

$$b_{Ci} = \frac{\sum_{j\in G,k\in L;j\neq k}\sum_{m=1}^{M(jk)} w_{jk}(m)c_{jk}(m)n^i(m)}{\sum_{j\in G,k\in L;j\neq k}\sum_{m=1}^{M(jk)} w_{jk}(m)c_{jk}(m)} \quad (5)$$

$d_{Li}$ 和 $d_{Ci}$ 为优化节点度数簇指标，$i$ 和 $j$ 为节点号；$l_{ij}$ 为节点 $i$ 与 $j$ 之间的电气距离（本文选取的电气距离为阻抗矩阵法）；$c_{ij}$ 为节点 $i$ 与 $j$ 之间的支路传输能力，即电网中两个节点最短路径所形成的线路集合中可承受的最小功率；$\alpha_{ij}$ 为电网节点邻接矩阵的元素，由于电网实际上是一个有向图，所以中的元素由 $i$ 流向 $j$ 为1，反之为–1，不相连为0。

优化节点介数簇由 $b_{Hi}$，$b_{Li}$，$b_{Ci}$ 组成；$k$ 为节点号；$G$ 为电网中所有发电节点的集合；$L$ 为电网中所有负荷节点的集合；$M(jk)$ 为发电节点 $j$ 与负荷节间 $k$ 的功率传输路径数；$w_{jk}(m)$ 为发电节点 $j$ 与负荷节点 $k$ 间的第 $m$ 条功率传输路径的权重系数，其数值等于这条线路上实际传输的功率；$n^i(m)$ 为发电节点 $j$ 与负荷节点间 $k$ 的第 $m$ 条功率传输路径是否经过节点 $i$ 的标识，若经过节点 $i$ 为1，否则为0；$d_{jk}(m)$ 为发电节点 $j$ 与负荷节点 $k$ 间的第条功率传输路径的电气距离；$c_{jk}(m)$ 为发电节点 $j$ 与负荷节点 $k$ 间的第 $m$ 条功率传输路径的传输能力。

电网节点 $i$ 的关键度指标是 $d_{Li}$，$d_{Ci}$，$b_{Hi}$，$b_{Li}$，$b_{Ci}$ 五个指标归一化后进行加权求和而得，用于辨识节点在电网结构和系统运行中的关键程度。

电网中某条支路的关键程度由优化支路介数簇加权求和得到，具体公式如下：

$$b_{He} = \frac{\sum_{j\in G,k\in L;j\neq k}\sum_{m=1}^{M(jk)} n^e(m)}{\sum_{j\in G,k\in L;j\neq k} M(jk)} \quad (6)$$

$$b_{Le} = \frac{\sum_{j\in G,k\in L;j\neq k}\sum_{m=1}^{M(jk)} w_{jk}(m)d_{jk}(m)n^e(m)}{\sum_{j\in G,k\in L;j\neq k} w_{jk}(m)d_{jk}(m)} \quad (7)$$

$$b_{Ce} = \frac{\sum_{j\in G,k\in L;j\neq k}\sum_{m=1}^{M(jk)} w_{jk}(m)c_{jk}(m)n^e(m)}{\sum_{j\in G,k\in L;j\neq k} w_{jk}(m)c_{jk}(m)} \quad (8)$$

$b_{He}$，$b_{Le}$，$b_{Ce}$ 分别为支路的优化支路指标簇；$e$ 为支路号；$n^e(m)$ 为发电节点与负荷节点间的第 $m$ 条功率传输路径是否经过支路的标识，若经过支路为1，否则为0。

电网节点的关键度指标是 $b_{He}$，$b_{Le}$，$b_{Ce}$ 三个指标归一化后进行加权求和而得，用于辨识支路在电网结构和系统运行中的关键程度。

## 2 基于改进图注意力网络的电力系统关键环节辨识

### 2.1 改进图注意力网络算法

图注意力网络是一种基于图结构数据的神经网络架构，它通过引入注意力机制到基于空间域的图神经网络中，与基于谱域的图卷积神经网络不同，图注意力网络不需要使用拉普拉斯等矩阵进行复杂的计算，而是通过一阶邻居节点的信息来更新节点特征，如图1所示。

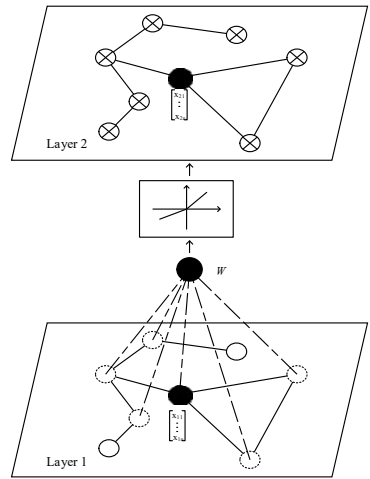

图 1 节点特征聚合示意图

Fig.1 Schematic of node feature aggregation

其注意力系数的计算公式如下：

$$\alpha_{ij} = \frac{\exp\{LeakyReLU[a^T(Wx_i \| Wx_j)]\}}{\sum_{m\in\theta}\exp\{LeakyReLU[a^T(Wx_i \| Wx_m)]\}} \quad (9)$$

式中：$\alpha_{ij}$ 为节点对于节点 $j$（$j\in\theta$，$\theta$ 为节点的邻居节点集合）之间的注意力系数；$a$ 为可学习的参数向量；$\|$ 为拼接操作；$W$ 是可学习的参数矩阵；$x$ 为节点特征向量；$LeakyReLU$ 是一种非线性函数。

之后，根据注意力系数对各节点信息进行加权聚合作为目标节点新的特征向量。

$$x_i = \sigma(\sum_{m \in \theta} \alpha_{ij} W x_j) \tag{10}$$

式中：$\sigma$ 为激活函数。此式为单头 GAT，多头注意力只需在式(10)的括号内头数相加取平均即可。

Brody 等人证明，原始的 GAT 的注意力函数是一种受限的静态的注意力[30]，将式 9 的计算顺序调整就可以得到动态注意力。如式 11 所示：

$$\alpha_{ij} = \frac{\exp\{a^T[LeakyReLU(Wx_i \| Wx_j)]\}}{\sum_{m \in \theta} \exp\{a^T[LeakyReLU(Wx_i \| Wx_m)]\}} \tag{11}$$

本文选取具有动态注意力的 GAT 为方法解决问题。即改进的 GAT。

### 2.1 基于 MGAT 的电力系统运行关键环节的辨识

电力系统本来就是一个错综复杂的网络，其网络拓扑加相应的数据即可构成能反映电力系统运行状态的图数据，具体来讲，其中发电站变电站负荷节点即为图数据的顶点，而输电线路即为该图数据的边，这样我们就可以把每个节点和支路都带有自己本身属性特征，本文选取特征量的维度如下：节点的特征为 5 维，支路的特征的维度为 6 维，见表1。

表 1 关键关节初始特征

Tab.1 Initial features of critical link

| 节点特征 | 支路特征 |
| --- | --- |
| 度数 | 度数 |
| 节点所带负荷有功功率 | 支路电抗 |
| 节点所带负荷无功功率 | 支路流过的有功功率 |
| 电压幅值 | 支路流过的无功功率 |
| 电压相角 | 支路负载率 |
|  | 支路损耗 |

在某一运行场景下对电网进行关键环节仿真，可以得到该状态下电网的关键环节，关键环节的仿真识别结果可以表示为 $m$ 和 $n$ 维的向量，$m$ 为节点数，$n$ 为支路数，作为关键环节辨识器的期望输出，两个向量中，1 代表关键线路，0 代表非关键线路。

利用 GAT 对辨识器的输入 $X$ 进行变换和聚合，得到聚合后属性，本文将聚合后的节点属性向量的维数设为 2，然后采用 Softmax 函数对节点特征进行归一化，获得节点或支路属于不同类别的概率。Softmax 函数公式如(12)：

$$p_i = \text{softmax}(y_i) = \begin{bmatrix} \dfrac{\exp(y_{1i})}{\exp(y_{1i})+\exp(y_{2i})} \\ \dfrac{\exp(y_{2i})}{\exp(y_{1i})+\exp(y_{2i})} \end{bmatrix} = \begin{bmatrix} p_{1i} \\ p_{2i} \end{bmatrix} \tag{12}$$

式中：$p_{1i}$ 和 $p_{2i}$ 分别代表节点 $i$ 属于关键环节和非关键环节的概率，将概率较大的类别作为节点所属的类别，$y_i$ 为节点 $i$ 得到的二维特征向量。

### 2.3 基于 MGAT 的电力系统关键环节辨识器的训练

#### 2.3.1 数据预处理

在训练模型之前，样本集的准备不可或缺，但是在 2.2 节得到的每个环节的特征每个维度之间的数值差异过大，比如节点度数和节点所带负荷的有功功率，他们之间的数量级差异很大，如果不处理就会出现训练集的 loss 正常减小，而验证集的 loss 没有明显的变化，模型基本没有泛化能力，不利于模型的训练，所以本文在训练之前，将具有同一属性的特征进行归一化处理，使得每个维度的特征都转化为一个范围的数值，这样既不会改变一组样本集上特征的差异，也会使得训练适应度更高。

#### 2.3.2 损失函数的选取

本文损失函数选取交叉熵损失函数(Cross Entropy Loss Function)，在神经网络训练过程中，由于 ReLU 等非线性激活函数的存在,神经网络经常出现梯度消失问题。采用交叉熵损失函数可以一定程度上解决这个问题。原因是交叉熵损失函数对于预测错误的样本赋予了较高的梯度,这样可以增加误差在反向传播中的传递,从而加速优化过程[33]。并且可以防止出现模型在训练集上表现良好、在测试集上表现较差的过拟合现象。采用交叉熵损失函数可以有效防止过拟合。这是因为交叉熵损失函数比均方差等传统损失函数更加敏感，更容易检测到噪声和异常值，从而避免模型过拟合，进一步提升模型泛化能力。其二分类公式为：

$$\begin{aligned} L &= \frac{1}{N}\sum_i L_i \\ &= \frac{1}{N}\sum_i -[y_i * \log(p_i) + (1-y_i) * \log(1-p_i)] \end{aligned} \tag{13}$$

式中：$y_i$ 表示样本的标签，正类为 1，负类为 0，$p_i$ 表示样本预测为正类的概率。

本文利用 Adam 优化器，对得到的节点特征向量进行更新优化，并使用 Dropout 方法防止出现过拟合现象。并采用多头注意力机制，头数为 2。

### 2.4 基于 MGAT 的电力系统关键环节的训练流程

基于 MGAT 算法的关键环节辨识模型的训练流程关键部分在于利用图数据和注意力机制完成对图节点特征的聚合。MGAT 分类训练过程的伪代码如下：

| Algorithm1：The classification training process of MGAT |
| --- |
| **Input**: training set $D\{(x^{(n)}, y^{(n)})\}$, maximum number of iterations $T$ |
| 1     Initial weight $\omega$ |

```
2      repeat
3        Randomly order the samples of the training set;
4        for  n = 1…N  do
5          Select a sample (x^(n), y^(n));
6          Calculate node and neighbor node attention coefficient;
7          Weighted aggregated node and neighbor node information;
8          Take the mean value of multiple attention node features;
9          Predict the categories according to formula(12);
10         If  ŷ_t ≠ y_t then
11           Calculate loss;
12           ω_{l+1} ← ω_l - γ[∂loss(ω)/∂(ω)];
13           l=l+1;
14         end
15         t=t+1;
16         if t=T then break;
17       end
18    until  t=T
Output: Optimal MGAT model
```

## 2.5 模型结果评估

采用混淆矩阵呈现模型辨识结果，矩阵组成如表2所示。

**表2 混淆矩阵组成**

Tab.2 Confusion matrix composition

| 预测结果 | 实际结果 | |
|---|---|---|
| | 是关键环节 | 非关键环节 |
| 是关键环节 | TP | FP |
| 非关键环节 | FN | TN |

通过准确率 $f_{acc}$ 精确率 $f_{pre}$、召回率 $f_{rca}$ 以及F1指标 $f_{F1}$ 全面评判测试结果，计算过程如下[34]：

$$f_{acc} = \frac{TP+TN}{TP+TN+FP+FN} \times 100\% \quad (14)$$

$$f_{pre} = \frac{TP}{TP+FP} \times 100\% \quad (15)$$

$$f_{rca} = \frac{TP}{TP+FN} \times 100\% \quad (16)$$

$$f_{F1} = \frac{2f_{pre} \cdot f_{rca}}{f_{pre}+f_{rca}} \times 100\% \quad (17)$$

## 3 算例与结果分析

本文采用改进的IEEE 6机30节点系统进行关键环节辨识方法的验证与评估。首先利用Matlab中的Matpower工具包仿真得到数据集，之后利用Python集成开发环境搭建基于GAT算法的脆弱性关键环节辨识模型。

### 3.1 改进的IEEE30节点系统

改进的IEEE30节点接线如图2所示，有30个节点，41条支路，本文在分析关键节点和关键支路时，选取关键环节得分排在前20%作为关键节点或关键支路，所以，该节点系统中，关键节点选取前6个，关键支路选取前8个。

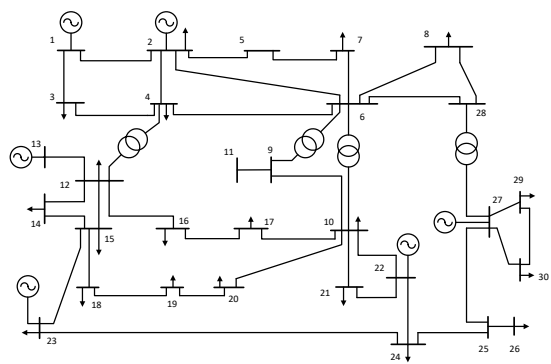

**图2 改进的IEEE 30节点接线图**

Fig.2 Improved IEEE 30-node wiring diagrams

#### 3.1.1 数据集的获取

为了可以真实模拟电网的运行，在电网拓扑不变的情况下，从发电机节点的出力、线路电抗和节点所带负荷方面更改参数来生成更多的断面数据，发电机节点的出力范围、线路电抗和节点所带负荷修改为原始情况的80%到120%，这样共迭代出11000个断面数据，除去不收敛的情况，共10350个断面数据。

#### 3.1.2 模型辨识结果

所建立的关键环节辨识样本集按照80%、10%、10%的比例被分为训练集、验证集和测试集。训练集用于调整权重矩阵辨识器参数；验证集用于检验模型训练过程中收敛、过拟合等情况，确定最优迭代次数等超参数；测试集用于测试最优辨识器的最终辨识能力。

文中通过数据预处理、设计交叉熵损失函数、加入多头注意力机制等方式对传统基于GAT算法的模型进行改进，得到本文所提关键环节辨识模型。本文模型在训练过程中迭代到20次时已经找到最好的结果，训练结果良好，关键节点和关键支路训练集和验证集损失率如图3和图4所示，关键节点和关键支路的模型的混淆矩阵如表4和表5所示：

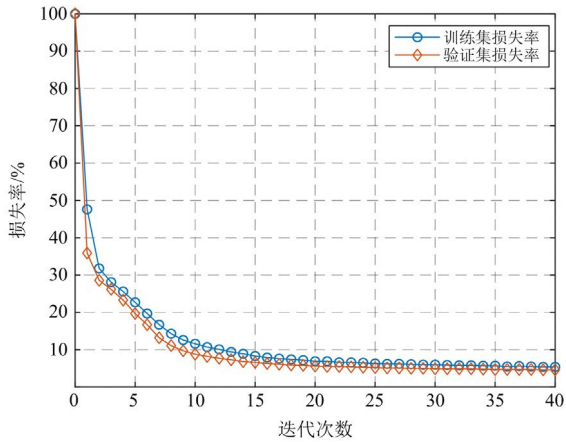

图 3 IEEE30 节点系统关键节点损失率情况

Fig.3 IEEE30 node system critical node loss rate situation

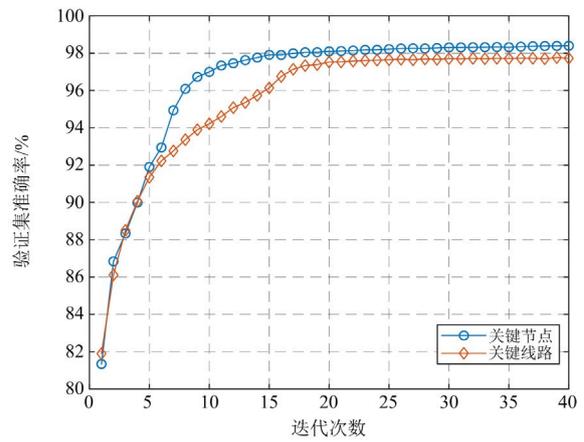

图 5 IEEE30 节点系统关键环节验证集损失率

Fig.4 IEEE30 Node System Critical Link Validation Set Loss Rates

表 3 IEEE30 节点系统关键节点辨识结果

Tab.3 IEEE30 node system critical nodes identification results

| 预测结果 | 实际结果 | |
|---|---|---|
| | 是关键节点 | 非关键节点 |
| 是关键节点 | 24531 | 261 |
| 非关键节点 | 223 | 5975 |

表 4 IEEE30 节点系统关键支路辨识结果

Tab.4 IEEE30 node system critical lines identification results

| 预测结果 | 实际结果 | |
|---|---|---|
| | 是关键支路 | 非关键支路 |
| 是关键支路 | 33732 | 357 |
| 非关键支路 | 602 | 7662 |

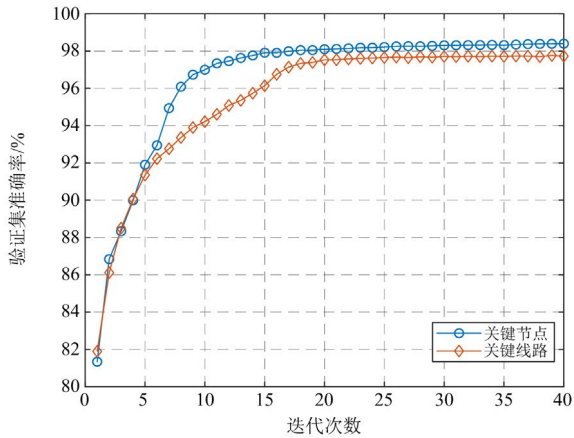

图 4 IEEE30 节点系统关键支路损失率情况

Fig.4 IEEE30 node system critical lines loss rate situation

经过多次测试，关键节点在测试集上的准确率可以达到 98.43%左右，效果优异；脆弱性关键线路在验证集上的准确率保持在 97.73%左右，效果优异，如图 5 所示。这正与上述两图相对应，在迭代 20 次左右正确率就满足基本要求。

综上，本文所提出的模型无论是在训练集还是验证集都得到的良好的辨识结果，辨识准确率满足一定的工程需要。

从表 3 我们可以算出本文设计出的关键线路的辨识模型的准确率、精确率、召回率及 F1 指标值分别为 98.3%、98.85%、98.85%、98.85%。从表 4 我们可以算出本文设计出的关键支路的辨识模型的准确率、精确率、召回率及 F1 指标值分别为 97.33%、99.56%、96.94%、98.23%。这些数据说明了本文辨识模型无论是在关键线路的辨识还是关键节点的辨识都表现出良好的辨识结果，符合精度上的要求。辨识模型的各项质变如表 5 所示。

表 5 IEEE30 节点系统辨识模型各项指标

Tab.5 IEEE30 node system identification model metrics

| | 关键节点 | 关键支路 |
|---|---|---|
| 准确率 | 98.43% | 97.73% |
| 精确率 | 98.95% | 98.95% |
| 召回率 | 99.10% | 98.24% |
| F1 指标 | 99.02% | 98.59% |

在模型评估出关键环节所消耗时间方面，本文选取 200 个场景作为测试来验证本文模型在快速方面的优势。本文所提方法所耗时由特征提取的时间和特征进入模型最后得到辨识结果的时间两部分组成。文献[31]的传统的基于仿真的方法在 30 节点系统得到关键节点和支路所耗时在 51s 左右（由于仿真生成的数据有随机性，所耗时间有微小误差）。本文方法在 30 节系统耗时 4s 左右（特征提取时间为 3s 左右，特征进入模型最后得到辨识结果的时间大约为 1s），比传统方法快 12 倍，更能适应如今多变的电力系统。

### 3.1.3 不同模型之间的比较

首先，为证明 GAT 算法在特征提取上的优势，在相同的样本库、模型框架等情况下，选择深度神经网络(Deep Neural Networks，DNN)、卷积神经网络(Convolutional Neural Networks，CNN)、循环神经网络(Recurrent Neural Network，RNN）和本文所用的 GCN 算法进行对比，各算法训练过程中验证集准确率如图所示，CNN、RNN 和 DNN 由于对权重初值要求高，结果不稳定，这里选择多次训练取平均值的方法来可视化其训练效果。

由图 6 和图 7 可知，在训练和测试中以 MGAT 为代表的图深度学习算法因在学习过程中计及了线路的电气特征和拓扑属性信息，所以在辨识准确性以及模型稳定性优于 DNN、CNN、RNN 为代表的仅考虑电气特征的深度学习算法。

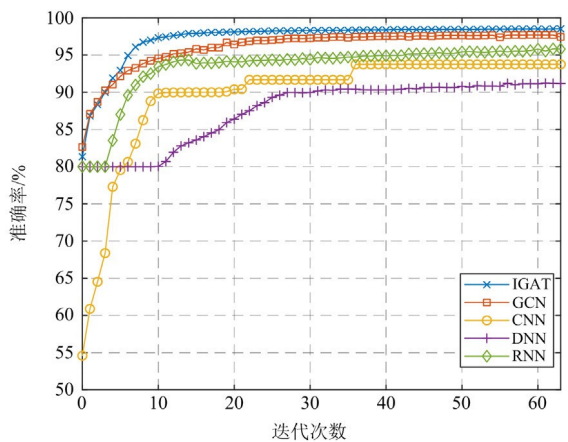

图 6 不同方法关键节点验证集准确率

Fig.6 Critical nodes validation set accuracy for different methods

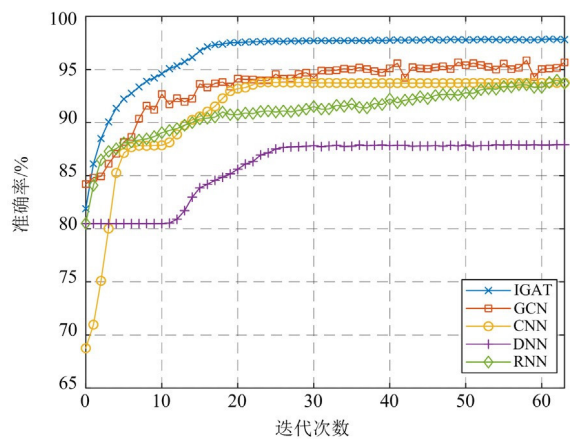

图 7 不同方法关键支路验证集准确率

Fig.7 Critical lines validation set accuracy for different methods

## 3.2 实际电网案例

本文选择东北某省的 220kV 的电网，该电网由 126 个节点（其中发电厂 28 个，98 个变电站/母线）、167 条输电线路构成，如图 8 所示。

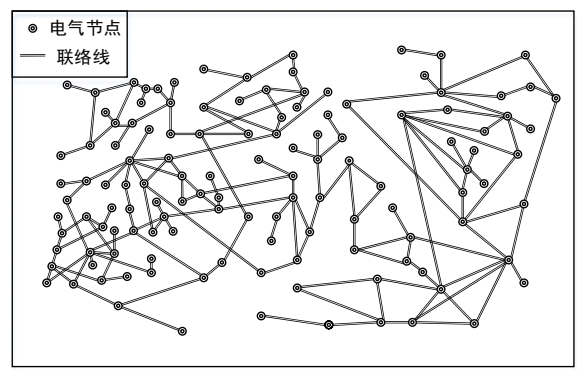

图 8 实际电网接线图

Fig.8 Actual grid wiring diagram

首先，利用 1.2 节方法评估各运行场景下关键环节，根据系统规模定义线路或节点为关键环节，其中，将评估指标值前 20 个节点定义为关键节点，将评估指标值前 30 个支路定义为关键支路；之后，根据实际运行数据和评估结果构建训练样本集。考虑到电网运行具有连续性、周期性等特点，选择具有代表性的运行场景。然后通过更改发电机节点出力和负荷节点所带负荷的方式进行扩充样本集，改变范围为原始情况的 80%到 120%。本文选择一天中连续的 4 个小时作为原始数据进行扩充，共生成 5000 组数据（其中原始 240 组，仿真生成 4760 组）。最后，对模型进行训练与测试，验证有效性。模型训练结果如图 9 所示。

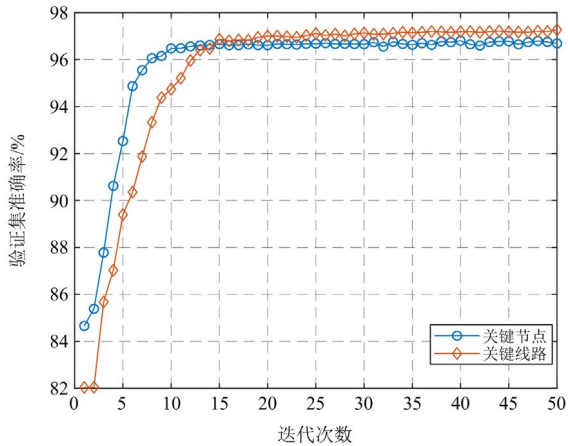

图 9 实际电网关键环节验证集准确率

Fig.9 Validation set accuracy of critical link of vulnerability in an actual power grid

经过多次测试，实际电网脆弱性关键环节的辨识模型的准确率可以达到 96.56%和 97.23%。模型的各项指标如表 6 所示。从表 6 可以看出，基于本文方法的实际电网辨识模型具有良好的性能，能够满足在线辨识的任务。

表 6 实际电网系统辨识模型各项指标

Table 5 Identification mode index of an actual power grid

%

| 指标 | 脆弱性关键节点 | 脆弱性关键线路 |
| --- | --- | --- |
| 准确率 | 96.56 | 97.23 |
| 精确率 | 97.77 | 98.12 |
| 召回率 | 98.17 | 98.50 |
| F1 指标 | 97.98 | 98.31 |

继续利用 200 个场景作为测试样本，利用传统评估方法得到脆弱性关键环节所需时间为 1200s 左右，本文方法所需时间为 97s 左右(特征提取时间 96s，模型辨识时间 1s)。这与上文仿真系统的结果一致，本文方法比传统方法评估脆弱性关键环节的时间大约快 12 倍。

## 4. 结论与展望

随着电网规模不断扩大以及复杂度逐渐提升，传统关键环节辨识方法在速度、准确性方面不能很好地满足当前电网运行实际需求。本文提出一种基于 MGAT 算法的关键环节辨识方法。该方法借助图深度学习的优势，利用 MGAT 算法深层次挖掘电网运行工况与关键线路之间的潜在联系，实现新运行工况下关键环节的快速、准确辨识。通过改进的 IEEE30 节点系统和实际电网算例验证了所提方法的有效性和可行性。

主要结论如下：

1）与传统的辨识方法相比，本文基于 MGAT 算法构造的辨识器通过离线训练后，对新运行状态下关键线路的辨识速度得到显著提升，更适合于关键线路辨识任务的在线应用；

2）与基于深度学习的辨识方法相比，采用图深度学习方法可同时对线路属性信息及其所处拓扑位置进行深层次学习，能获得更完备的辨识信息，辨识结果的准确率更高。

值得说明的是，本文在所应用的关键环节模型中主要考虑了系统的拓扑关系，而如何在本文模型基础上改进使其更能适应如今新能源并网情况下的关键线路的辨识；关键环节评估后对于连锁故障的影响；新的人工智能技术的应用等，将是未来需要进一步研究的问题。